# Outflanking and Securely Using the PIN/TAN–System

A. Wiesmaier, M. Fischer, M. Lippert, J. Buchmann
Department of Computer Science
Technische Universität Darmstadt
D-64283 Darmstadt, Germany

*Abstract*— The PIN/TAN–system is an authentication and authorization scheme used in e–business. Like other similar schemes it is successfully attacked by criminals. After shortly classifying the various kinds of attacks we accomplish malicious code attacks on real World Wide Web transaction systems. In doing so we find that it is really easy to outflank these systems. This is even supported by the users' behavior. We give a few simple behavior rules to improve this situation. But their impact is limited. Also the providers support the attacks by having implementation flaws in their installations. Finally we show that the PIN/TAN–system is not suitable for usage in highly secure applications.

## I. INTRODUCTION

### A. Classification and related work

Since a couple of years we have experienced growing usage of internet based services in fields like e–banking, e–commerce and e–government [1]. As the internet and its utilization by the population are still growing, those services will become a more essential part of our life [2].

The increasing number of attacks against online services shows that insecure internet transactions are a heavy threat to the society. The most famous attacks are the so–called "phishing"–attacks. The Anti–Phishing Working Group counted over 1500 active phishing sites in November 2004 [3]. In June 2004 Cowley [4] reported about a key logging Trojan which captures passwords. In August 2004 Mikx [5] demonstrated a proof of concept for a MIM (**M**an **I**n the **M**iddle) attack capable of stealing passwords. The Gartner Group [6] estimated that in 2003 nearly 2 million Americans have had their checking accounts raided by criminals. Schneier [7] predicts that this number will be even higher for 2004.

### B. Contribution and outline

The scope of this paper is the PIN/TAN–system (**P**ersonal **I**dentification **N**umber / **T**rans**A**ction **N**umber). It is a common scheme for securing sensible transactions in Europe [8]. As mentioned above this system is heavily threatened by various attacks. We investigate why it is vulnerable and what can be done to improve the situation.

As it is very widespread, and therefore easy to examine, we initially focus on web based internet banking as an example online transaction platform. Many European online banks support the PIN/TAN–system and it is well accepted by the banking customers [9]. We next focus on spyware attacks. Later in this paper we will generalize our results and thereby see that the PIN/TAN–system is inherently insecure, independent from the transaction platform or the kind of attack.

In order to reach our goals we implemented and accomplished real spyware attacks on real online banking accounts. Thereby we learned about the PIN/TAN–system, which problems an attacker has to solve and how to hamper or prevent attacks. We started this project in 2001 with the master thesis of Fischer [10] where we attacked a demo online account provided by a real online bank. In 2002 we refined our tools and finally attacked real online accounts successfully. Due to legal issues we had to keep our results secret. The thereafter emerged real world attacks prove our precognition of the threats in insecure online transactions. Now we are able to present our results.

This paper splits into two parts. The next three Sections are quite practical. In Section II we describe how the PIN/TAN–system works. This is the basis for the remain of the paper. For gaining an overview we shortly classify the various kinds of attacks in Section III. After that, in Section IV, we explain how we implemented our attack.

The second part is more analytical. We elaborate on new insights we derived from accomplishing the attacks. In Section V we see that the customers' behavior supports being attacked. The typical (mis)behavior is shown. We investigate how this situation can be improved by better behavior in Section VI. Section VII shows that also the providers support the attacks by having massive implementation flaws in their systems. Those flaws are discussed. The fact that the PIN/TAN–system is inherently insecure is shown in Section VIII. Section IX concludes the paper.

## II. PIN/TAN HOME BANKING

At first we examined how customers experience web banking with PIN and TAN. Therefore we investigated a couple of internet banking providers. We encountered that all the examined online banking systems work in very similar ways.

We solely found client–server–systems. They split into HTTP (**H**yper**T**ext **T**ransfer **P**rotocol) based and applet based systems. We give a simplified overview of both systems.

The HTTP based system, shown in Figure 1, is the simpler one. The users connect to the bank using their web browsers. The customer's web browser and the bank's web server communicate using HTTP. All content is encoded in HTML (**H**yper**T**ext **M**arkup **L**anguage) and JavaScript and is displayed directly by the browser.

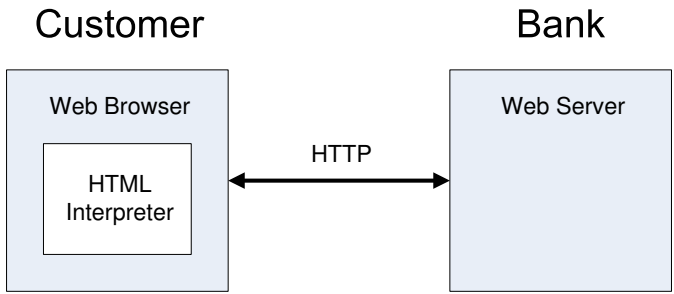

Fig. 1
HTTP BASED ONLINE BANKING

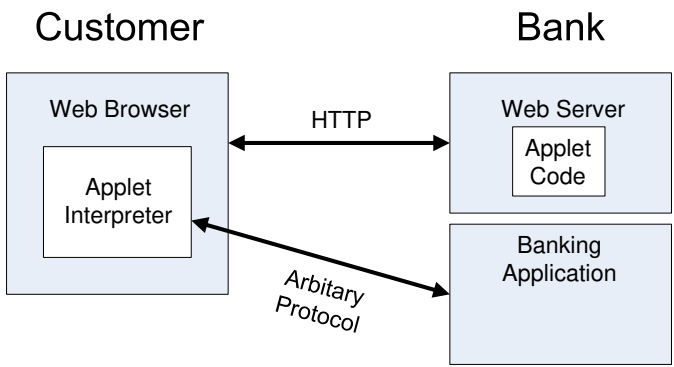

Fig. 2
APPLET BASED ONLINE BANKING

The applet based system, shown in Figure 2, is more flexible. The customers again connect to the bank using their web browsers and HTML. But then they automatically receive an applet which is their banking client. The communication between the customer's applet and the bank's server now can be based on any standard or proprietary protocol. The content can be coded in any format and is displayed by the applet.

In both versions the bank is authenticated via SSL/TLS (**S**ecure **S**ocket **L**ayer / **T**ransport **L**ayer **S**ecurity). This mechanism is also used to achieve the confidentiality and integrity of the data transferred in both directions. The purpose of the PIN/TAN–system is to authenticate the customers and to authorize the access to their accounts. The customer receives an ID (**ID**entifier), the PIN and a list of TANs from the bank in an out of band way (e.g. postal service). While it is possible to change the PIN, the ID and the TANs are unchangeable. While the ID might be public the PIN and the TANs are secret and only known to the respective customer and the bank.

In order to log in the user fills in the ID (e.g. a customer number or an account number) and the PIN into the respective form fields on the screen. The ID identifies the user. It tells the bank who wants to start doing transactions. The PIN is used to authenticate the user. It "proves" the identity of the user to the system. It is used like a usual password. A user who is logged in can perform all "read" operations like viewing the current balance or listing the standing orders. The PIN is valid multiple times until it is actively changed by the user. The ID stays valid until the account is canceled.

Each "write" operation (like transferring money or setting up a standing order) has to be authorized by a fresh TAN. After filling in the transaction data the user fills in the next valid TAN into the respective form field on the screen. The transaction data tells the system what to do. The valid TAN authorizes the access and indicates the "transaction release". Each TAN can be used only once. Thus they are used like one time passwords. Using a TAN invalidates it (and usually all its predecessors) for all further operations. The bank will send a new TAN list before the user runs short on remaining valid TANs.

Some online banks additionally use a BEN (**BE**stätigungs**N**ummer[1]). This is a check number which is returned by the online bank after a successful transaction. Each BEN is assigned to a respective TAN. The customer can easily audit this number[2] in order to ascertain that the transaction ran correctly.

The users's text input is done via the keyboard. In some cases it is additionally possible to use virtual keyboards, which are displayed on the screen and operated with the mouse. The positioning of the curser can be done with the mouse or the "Tabulator"–Key.

### III. CLASSIFICATION OF ATTACKS

We examine the PIN/TAN–system for its vulnerabilities. Therefor we have to deal with various attacks. This section gives a short overview of the different kinds of attacks we deal with. As it is out of the scope of this paper we do not mention attacks that are based upon compromised service providers.

#### A. Man in the middle

A MIM attack is a scheme where the attacker is able to read, modify or disconnect at will the communication between the customer and the online bank. The attacked parties do not know that they do not have a direct connection, and that the attacker sits in between. The MIM software may be located at the customer's computer or anywhere in the network.

#### B. Spyware

Spyware attacks eavesdrop on the communication between the customer and the service provider. The spy can read the communication, but is not able to modify the messages. In some cases the spy is able to disconnect the link between the parties. The spy may be on the user's machine or in the network.

---

[1] German: acknowledge number
[2] e.g. it is printed beside the TAN

## C. Phishing

Phishing attacks use spoofed emails or fraudulent websites to obtain passwords and other secrets. The victims are convinced that they communicate with their legal service provider and reveal their secrets to the criminals. The distinctive attribute of phishing attacks is that the criminal just claims to be the user's legal service provider. In opposite to the other attacks phishing does not base upon a real communication between the attacked parties. The service provider isn't involved at all.

## IV. ACCOMPLISHED ATTACK

The issue of this Section is the question how to outflank the PIN/TAN–system and whether this is easy.

Our approach to investigate this was to implement our own attack on the system. Thereby we learned about the PIN/TAN–system, how to accomplish an attack and how to hamper or even prevent attacks. We decided to keep the software very simple. This is enough for gaining experience and finally have a proof of concept. We describe our attack in this Section.

As the result we can see that it is very easy to outflank the PIN/TAN–system. We did not need any special hacker skills to arrange this attack. Some practical experience with online banking, some basic knowledge in programming and a little logical thinking are enough to outflank the PIN/TAN–system. The shown attack was successfully accomplished on real world online accounts that we own.

### A. The idea

The sketch for the attack is as follows: Infect the legal user's system with a computer virus or a Trojan containing a spy. This spy lurks hidden in the background and eavesdrops on the keyboard and mouse to obtain the ID, the PIN and the TAN. If all information is obtained the spy closes the web browser. This is done before the TAN is sent to the bank server. Thus the TAN stays valid. Then the stolen information is used to transfer money from the victim's account to the raider's account.

### B. The target

We chose the operating system Windows together with the web browser Internet Explorer as the platform to be attacked. This combination qualifies as a good basis for two reasons. Firstly, this configuration is extremely common [11], [12]. This leads to a very big pool of potential victims. Secondly, both programs had, have and probably will have a lot of security gaps [7], [13], which even aggravate in their combination. Those can be exploited to accomplish the attack.

We decided to attack a bank offering a demo online account on its web site. This demo actually is for showing potential customers how easy and comfortable online banking is. But this demo account can be misused for practicing attacks on online accounts — absolutely anonymous.

After attacking our first target bank we also attacked some other banks. We did this to show that it is the PIN/TAN–system self which is insecure, independent of who utilizes it. In fact we had to make only slight modifications to our tools and also succeeded in those attacks.

### C. The spy

To learn a lot about building spies, we build the software from scratch. This was relatively easy as the regarding Windows and Internet Explorer hooks are well documented. But we also investigated the possibility of utilizing downloadable hacker software. We chose a famous "remote administration software". The respective sources can be downloaded for free and have the spying features already built in. We just had to add the functionality for telling apart IDs, PINs and TANs from each other and from other input.

Once started, our spy waits in the background for the input of the desired information. When the victim starts his online banking activity, the spy collects all information needed for executing a bank transaction by eavesdropping on keyboard and mouse events. Having all desired data the spy terminates the browser before it has time to send the TAN (which is the last data) to the bank. According to Dvorak [14], the average Windows installation crashes three times a month. Thus, the low–brow online banking user will not be too amazed. As we will see in the following even a distrustful user might have no chance to cancel or undo the attacker's transaction.

Now the spy holds the full account information and the authentication / authorization data (including a valid TAN). The information is immediately sent to a central server which uses it to directly remit some money to the attacker's account. This automatic transaction requires much less time than the victim needs to reopen the web browser, surf to the online bank and log in again. Thus we can be sure to use the valid TAN before the victim uses it again and invalidates it. If the victim reuses the TAN he will be told that it is invalid because it was already used. But as he has entered this TAN just before the web browser crashed, he will probably not be too surprised. The user will either think the transaction has succeeded,[3] or he will use the next TAN to execute the transaction.[4]

### D. The traitor

Again we tried to learn a lot, and built the Trojan from scratch. In order to give the spy software the capability to unnoticeably be distributed widely, it was hidden within a useful tool, a currency converter. If the tool is installed, the spy is — invisible to the user — installed also and is automatically started each time Windows is booted.

We also investigated the usage of out of the box hacker software to directly infect the potential victims' computers. Everyone is able to download ready compiled binaries which enable the execution of arbitrary code on unsecured external computers. A very famous exponent can be found easily in the internet. It is an open source project declared as penetration testing software. Its architecture allows it to always exploit the newest security leaks. The community always programs the newest exploitation code which then can easily be plugged into the existing installation.

---

[3] if BEN is utilized: and the BEN was lost due to the browser crash
[4] if BEN is utilized: and receive a correct BEN from his bank

### E. The hidden

The wide distribution of the spy respectively the holding of many valid transaction data records provides the attacker, beside the chance to deprive many online banking customers, the possibility to cover his tracks. The raider could, instead of transferring the grifted money directly to his account, remit the amounts some times to and fro between the disclosed accounts. Thus, he has enough time to break away before the trace can be backtracked to him. Another possibility is to donate parts of the money to foreign people in order to hide in a crowd of involuntarily enriched people. In order to be able to easily apply those and similar obfuscation tactics, the stolen account information is not used directly by the spy but is send to a central server. This server executes the account transactions due to actual tactic and current state. A more sophisticated way of doing this would be to post the transaction data encrypted or steganographed to one or more public message boards or news groups. Then the obfuscation control could be implemented within the spy. Thus a central server — which implicates a risk to be discovered early — can be omitted.

## V. Users' behavior

The issue of this Section is the question how far the online banking user is to blame for our success in attacking the PIN/TAN–system.

Our approach to investigate this was to examine the typical user behavior. We did this especially anent our attack. This Section lists the shortcomings in user behavior we found.

The result is that we found some points in user behavior which helps the eavesdropper in tapping the data. We further see that this behavior does not weaken the PIN/TAN–System in special. It weakens all transaction security systems. We found some real misbehavior concerning the security (levity, carelessness). But we also found some behavior that indeed supported our attack, but not really can be seen as misbehavior (normal input behavior, mainstreaming).

### A. Levity

The levity of the end user is a major shortcoming. Without having the feasibility to infect end systems with malicious code the shown attack and similar ones are not possible. It's a matter of common knowledge that the infection of unsecured systems with viruses or worms is very easy. This is also true for systems where the passwords are weak or even not set. The worst thing a user can do is to download unknown code from the internet or from email attachments and execute it. This is the surest way to infect the own system with malicious code.

### B. Normal input behavior

A very big aid for programming the spy was the existence of something like a predictable "normal input behavior". We hereby mean that a user usually fills in the data into a form field in its "natural sequence". If, for example, the user has to fill in his TAN, let's say "123456", he will type at first the '1', then the '2', after that the '3' and so on. Finally the input is terminated by a non numerical character. In the majority of cases this is the "Enter"–Key, in fewer cases the "Tabulator"–Key or a mouse click. This is true as long as the user does not accidentally mistype. As an attacker is comfortable with having a high probability for getting valid data, the spy does not have to read the content of the various fields which is hard to implement. It is satisfactory to eavesdrop on the keyboard input stream and the mouse position and clicks. By doing this one can easily grep the substrings which consists of numbers only. Having these number strings you just have to tell apart IDs, PINs, TANs and the various account numbers from each other. Those things emerged as very easy to implement.

### C. Carelessness

We talked with various people using online banking and experienced the following:

- Most people do not care about the security of online banking at all, until they are asked about it.
- Most people have the opinion that the PIN/TAN–system is secure just because the online banks utilize it.
- Most of the people think the bank clerk is an expert for online banking security.
- Most people did not change to more secure systems after being told about the weakness of the PIN/TAN–system.

Surely our survey is not fully representative, but at any rate it shows the tendencies. The average online banking user is to blame for having nearly blind trust in the "institution bank". They do not show the necessary responsibility for their own concerns.

### D. Mainstreaming

As already mentioned in section IV-B it is more efficient for an attacker to concentrate on popular systems. The more a system is used the more potential victims exist. This means on the other hand the more popular your system is the higher is the probability to become a victim. Thus, using a popular banking client on a popular operating system for contacting a popular online bank is, under the aspect of security, a bad idea.

## VI. Improved behavior

The issue of this Section is the question whether attacks can be hampered or prevented by the users on their own.

Our approach to investigate this was to improve the behavior we discussed in Section V. We consider only measures which can be applied by the willing customer on his own initiative, without being dependent on the cooperation of the online bank. This Section proposes rules for an improved user behavior.

As the result we see that there are some measures which can be applied by the customer to improve the situation. But we see also that those measures are trivial, uncomfortable or strange. In addition their impact is limited. They only prevent very simple attacks. More sophisticated attacks are barely hampered. Thus there must be other faults in the system which enable attacking the system.

## A. Prudence

This Section addresses the shortcomings explained in Section V-A. As a basic principle each computer shall be secured properly, independent from the usage purpose. If a computer deals with sensitive data — as it is the case when using it for online banking — this will be a mandatory task for the prudent user. There are a couple of (easy) things to be done to hamper an infection with malicious code:

- Use anti–spyware, anti–virus, and firewall software.
- Keep the above mentioned, the operating system and your other applications up–to–date by always having the latest security updates and patches installed.
- Do not execute unknown code downloaded from the internet or received by email.
- Use strong passwords. Be aware that malicious code might try to guess your passwords.
- Stay up–to–date in IT–security concerns. Subscribe to an IT–security newsletter or frequently visit a respective website.

Naturally this does not protect a system to 100%. But it increases the security noticeably. It prevents infections which use simple tricks or are based on older security holes. Infecting well secured systems is very hard and requires a high amount of special know how and criminal energy. Similar recommendations are made since a couple of years e.g. [15] or [16].

## B. Confusion

To improve the behavior described in Section V-B do not enter your data in the "natural" order. This means use some or all of the following tricks for confusing a potential spy.

- Do not fill the form from the left upper corner to the right lower corner. Choose the sequence of filling in the individual fields randomly.
- Do not fill the individual fields from the left to the right. Fill a field just partly, jump to another filed, fill it partly, jump back, complete this field, and so on.
- Insert a mistype from time to time and make intensive use of the "Del"– and "Backspace"–Key
- Change the position of the cursor with various methods. Use the "Tab"– and "Backtab"–Key, the mouse and the cursor keys randomly.
- Use the "Copy and Paste"–Mechanism for filling the fields.

Spyware which is not confused by those measures must be able to read and interpret individual form fields. Building such software is much harder than just installing a "blind" eavesdropper on the input devices. Thus, although these hints do not fend off all possible versions of spyware, they at least protect against a part of them. But we see also that those recommendations are strange and uncomfortable.

## C. Carefulness

This Section addresses the shortcomings mentioned in Section V-C. The simple rule here is: "Care about security." Be aware that a bank is in first degree a commercial organization with the primary goal to draw profit. Online banking is much cheaper for the bank than having a lot of branches full of clerks. If the people pay more attention to security while choosing their online banks, the banks are forced to introduce more secure systems.

## D. Individualization

For avoiding the misconduct figured out in Section V-D one should not use software used by many others. Choose an operating system, a service provider and client software which are not very widespread. Certainly individualization does not protect your system but it decreases the probability to be attacked noticeably. As noted in Section IV-B an attacker probably will concentrate on very popular systems to gain a big amount of possible victims.

## VII. IMPLEMENTATION FLAWS

In the preceding Section we saw that the users have little to no chance to increase the security in dealing with the PIN/TAN–system. The issue of this Section is the question whether the banks make mistakes in dealing with the PIN/TAN–System.

Our approach to investigate this was to search for flaws in implementing the PIN/TAN–system. The banks surely would not reveal their implementations to us. Thus we focused on those details which can be revealed through the normal usage of the system. For this purpose we opened some accounts with German banks offering online services. This Section lists the implementation flaws we found there.

The Result is that there are some heavy implementation flaws which really simplify attacks like the one we showed here. It revealed that we exploited some of the found implementation flaws in our attack. All flaws are not special PIN/TAN–system implementation errors. They weaken all online transaction security systems. They can be fixed without making changes to the PIN/TAN–system. It is on the providers to do so. We have a work in progress [17] where we propose suitable measures.

### A. Aborting transactions

A big flaw, if not the major flaw, is aborting the transaction if the TAN does not arrive at the service–provider. Whatever a user sends to the service provider will be handled as never arrived if the authorizing TAN does not arrive. This means that the service provider will accept a new transaction from this user without any further measurements. It does not matter that the previous transaction was aborted in an exceptional way. Thus an attacker can easily obtain a valid and usable TAN by allowing all user data to be passed over to the service provider and just intercepting the TAN. This is exactly what we did in our attack.

### B. Concurrent sessions

Another failure was the admission of concurrent sessions of the same user. It does not matter that a user is already logged in, it is possible to log in again, even from a different site. It does not matter that the user is about to commit a transaction from site A, the same user can start and commit a concurrent transaction from site B. Even if this might be comfortable, it is probably not necessary and for sure it is dangerous. This means a local spy does not have to intercept the TAN. If the spy has a concurrent session open, it is sufficient to just eavesdrop on the TAN and use it at first. This should be no problem as an automatic detection and usage of a TAN is for sure faster than the human break between typing in the last digit of the TAN and pressing the "Enter" button. Another effect is that intercepting spies do not have to care for session timeouts or similar things. They can log in as soon as all necessary data is available. This is the way we exploited this gap.

*C. Distinguishability*

The easy distinguishability of the various user inputs is a further fault. It is very easy to tell the ID, the PIN and the TAN apart from each other and the remaining input. Firstly, they appear in a given order. This means an attacker knows which information comes next. Secondly, they differ in their length and the allowed characters. This means an attacker knows where a special piece of information ends and where the next starts. Thus, an attacker does not have to read and interpret the various form fields. It is sufficient to eavesdrop on the data stream to grip the desired information. Our spy works this way.

*D. HTML simplicity*

The next fault is the very simple and straight forward HTML code used by the providers. The sequence of the web sites, their respective URL (**U**niform **R**esource **L**ocator) and the names of the POST data variables are always the same. We experienced only once a minimal change after one provider updated his web appearance. This makes it very easy to implement robots which execute automatic transactions with stolen access data. Such a robot is part of our attack.

*E. Insecure out of band mechanism*

An interesting flaw is the usage of insecure out of band mechanisms to exchange the access information. The ID, the PIN and the lists of TANs are sent with (paper) mail in separate letters. Normally with a few days between them (but one time they came at the same day). The providers do not tell the customers when the letters should arrive. There is no direct feedback whether the mail has reached the addressee. Thus, stealing the access data out of the customers (physical) mailboxes is an easy way for obtaining them. We did not exploit this flaw, since it is not suitable for electronic mass attacks.

## VIII. Enervations of the PIN/TAN–system

All failings we found up to now weaken the PIN/TAN–system. But they are not special PIN/TAN–system failings. Further more up to now we focused on spyware attacks. The issue of this Section is whether the PIN/TAN–system has inherent flaws and whether those regard all kinds of attacks.

Our approach to investigate this was to take a detailed and somewhat theoretical look at the PIN/TAN–system. We did not focus on our initial scenario but investigated the raw and abstract form of a PIN/TAN–system. This Section lists all weaknesses we found.

The Result is that the PIN/TAN–system is inherently weak because it has heavy flaws. This is the reason why it is not possible to protect the original system against the mentioned attacks. Thus the PIN/TAN–system is unsuitable for securing online transactions. It revealed that we exploited some of the found inherent flaws. It is on the providers to utilize better authentication / authorization schemes. We have a work in progress [17] where we propose new schemes.

*A. Impersonation*

From the theoretical view the PIN/TAN–system provides authentication and authorization by knowledge. Who ever knows the ID and the PIN is able to log in. The provider will accept the holder as the respective legal user. Who in addition knows valid TANs is able to perform online transactions. The provider will accept each transaction as authorized by the respective legal user. And this without any additional arrangements and from everywhere in the world. This means obtaining the valid ID, PIN and TANs is enough to be able to completely impersonate the respective legal user.

*B. Clear text*

The ID, the PIN and the TANs are typed in unaltered, complete and in clear text into the computer. Thus valid authentication and authorization data can be obtained by just somehow "stealing" the victims input. It is not necessary to obtain the original PIN letter or the original list of TANs.

*C. Replay*

The ID stays valid as long as the online account exists. The PIN stays valid until it is actively changed, regardless if it is used or not. Thus the login event is vulnerable to replay attacks. Who ever is able to eavesdrop on a successful login is (with a high probability) able to login as the victim.

*D. Relationship*

There is no conjunction between a valid TAN and the thereof authorized transaction. Each valid TAN can be used for any transaction. This means intercepting a valid TAN which is designated by the victim to authorize transaction A can be used by the attacker to authorize an absolutely different transaction B.

*E. Dependence*

The PIN/TAN–system has no arrangements to detect (or prevent) the eavesdropping, the interception or the exchange of the users input. This must be accomplished by another mechanism which has to be used in conjunction with the PIN/TAN–system. If this mechanism is weak or does not fit perfectly, this presents a gap which can be exploited by an attacker.

*F. Receipt*

The PIN/TAN–system does not offer revisable receipts to the user. The user has no chance to prove to a third party that he instructed a certain transaction to the service provider. As the user and the service provider both know the valid BENs they can not be used by the user to prove this. Furthermore the used BEN depends only on the used TAN and not on the actual transaction. In case of discrepancies between the user and the service provider those must be solved by arrangements in the contract or by goodwill.

*G. Repudiation*

The PIN/TAN–system does not offer revisable orders to the bank. The bank has no chance to prove to a third party that the user has instructed a certain transaction. As the user and the service provider both know the valid TANs they can not be used by the bank to prove this. Furthermore the used TAN depends only on their sequence in the list of TANs and not on the actual transaction. In case of discrepancies between the user and the service provider those must be solved by arrangements in the contract or by goodwill.

## IX. CONCLUSION

We showed that it is easy to outflank the PIN/TAN–system. We proved this by accomplishing attacks on real online transaction systems. It emerged that the described attack is particularly dangerous due to the fact that it is not necessary to have special expertise in transaction protocols or hidden security leaks. Some basic knowledge in programming and a little motivation are enough to be able to deprecate other peoples' online accounts. We also saw that having a big set of transaction data helps the attacker to hide his traces and improves his chances to escape.

It further emerged that the customers support being attacked by their behaving. This behavior was identified and explained. We gave some proposals how the customer can increase the security without any assistance of the online bank. Even though these recommendations slightly increase the security, we saw that they are trivial, uncomfortable or strange. In addition they can not lift the level of security to a really high level.

Next we saw that even the providers support the attacks. They have defects in their implementations which ease attacking their systems. Those flaws are not related to the utilization of the PIN/TAN–system. But in combination with the PIN/TAN–system they worsen the situation noticeably. It is the providers' task to patch these holes.

Finally we listed and explained the enervations of the PIN/TAN–system. It exposed that the weak points are not caused by the concrete attacked platform or the actual kind of attack. They are inherent in the PIN/TAN–system. Thus, it is not possible to utilize it in a really high secure internet transaction system without further measurements. The providers must introduce more secure systems.

## REFERENCES

[1] J. Carlson, K. Furst, W. Lang, and D. Nolle, "Internet Banking: Market Developments and Regulatory Issues," May 2001, http://www.occ.treas.gov/netbank/SGEC2000.pdf (23 Sep. 2004).
[2] Ipsos News Center, "Retail Banks Threatened By Online Bill Payment Competition," November 2004, http://www.ipsos-na.com/news/pressrelease.cfm?id=2454 (18 Nov. 2004).
[3] Anti-Phishing Working Group, "Phishing Activity Trends Report," Nov 2004, http://www.antiphishing.org/APWG%20Phishing%20Activity%20Report%20-%20Nove Dec. 2004).
[4] Stacy Cowley, "Net Threat Targets Banks," June 2004, http://www.pcworld.com/news/article/0%2Caid%2C116761%2C00.asp (15 Dec. 2004).
[5] Mikx, "Alpha Phishing," Full–Disclosure Mailing List, August 2004, http://lists.netsys.com/pipermail/full-disclosure/2004-August/025744.html (03 Mar. 2005).
[6] Bob Sullivan, "Survey: 2 million bank accounts robbed," Website, 2004, http://msnbc.msn.com/id/5184077/ (13 Dec. 2004).
[7] B. Schneier, "Customers, Passwords, and Web Sites," *Security & Privacy Magazine, IEEE*, vol. 02, no. 5, p. 88, Sep-Oct 2004, http://www.schneier.com/essay-048.html (17 Dec. 2004).
[8] Celent Communications, "Online Banking Security in Europe: Balancing security, ease of use, culture and cost," July 2003, http://www.celent.com/PressReleases/20030709/OnlineBankingSec.htm (23 Sep. 2004).
[9] C. Tabbert, "HBCI im Wettbewerb der Homebanking–Standards," *Banking and Information Technology*, vol. 1, no. 1, pp. 32–40, March 2000, http://pc50461.uni-regensburg.de/NR/rdonlyres/AE54A0DC-130E-4E5B-B54A-98057I (22 Dec. 2004) in German.
[10] M. Fischer, "Angriff auf ein PIN/TAN–gestüztes Online–Banking–Verfahren," Master's thesis, Technische Universität Darmstadt, Germany, July 2002, http://www.informatik.tu-darmstadt.de/ftp/pub/TI/reports/Mikefish.diplom.pdf (23 April 2004) in German.
[11] J. Markoff and S. Hansell, "Apple Changes Course With Low–Priced mac," *The New York Times*, January 2005, http://www.nytimes.com/2005/01/12/technology/12apple.html?ex=1109998800&en=af2 (03 Mar. 2005).
[12] Mittel Deutscher Rundfunk, "Sicherheit beim Online–Banking," Website, July 2004, http://www.mdr.de/mdr-info/verbrauchertipps/1489307.html (03 Mar. 2005) in German.
[13] J. Leyden, "IE exploits top web security threat list," The Register, November 2004, http://www.theregister.co.uk/2004/11/02/web_security_survey_scansafe/ (03 Mar. 2005).
[14] J. Dvorak, "Magic Number: 30 Billion," PC Magazine, April 2003, http://www.pcmag.com/article2/0,4149,1210067,00.asp (03 Mar. 2005).
[15] B. Schneier, "Crypto–Gram December 15, 2004: Safe Personal Computing," December 2004, http://www.schneier.com/crypto-gram-0412.html#10 (17 Dec. 2004).
[16] Association of German Banks, "Online Banking Security," Website, July 2004, http://www.bankenverband.de/pic/artikelpic/072004/0407_Online_Sicherheit_en.pdf (03 Mar. 2005).
[17] A. Wiesmaier, M. Lippert, and J. Buchmann, "Improving the PIN/TAN–System," in *to be published*.